\documentclass{article}

\usepackage{spconf}
\usepackage{amsmath,amssymb}
\usepackage{graphicx}
\usepackage{booktabs}
\usepackage{array}
\usepackage{cite}
\usepackage[hidelinks]{hyperref}
\usepackage{multirow}

\title{PAG-AGENT: ZERO-SHOT PERSONALIZED MUSIC ALBUM GENERATION VIA AESTHETIC PROFILE DISTILLATION}

\twoauthors
  {Yifeng Mao, Ziye Wang, Yutong Shen}
  {Beijing University of Technology}
  {Feiran Liu, Wenjuan Han\sthanks{Corresponding author: wjhan@bjtu.edu.cn}}
  {Beijing Jiaotong University\\
   Email: wjhan@bjtu.edu.cn}

\begin{document}
\ninept
\maketitle

\begin{abstract}
Large Music Models (LMMs) can synthesize high-fidelity audio and support increasingly flexible prompt control, yet modeling the subjective preferences of an individual listener remains underexplored. We present PAG-Agent, a zero-shot multimodal framework that uses an explicit Personal Aesthetic Profile to bridge user preference inference and music generation control. A multimodal LLM distills favorite-song audio and contextual user information into this structured representation, which a planning LLM translates into generation instructions for existing music synthesizers. We further introduce PA-Eval, which evaluates personalization through the Aesthetic Consistency Score (ACS) for target-user alignment and the Aesthetic Specificity Index (ASI) for cross-user specificity. Experiments with Personal-100 on multiple MLLM and music-generation backbones show consistent improvements over generic and direct conditioning. A 50-participant follow-up study further shows a correlation \(r=0.64\) between ACS and subjective preference.
\end{abstract}


\section{Introduction}
\label{sec:intro}


Text-to-music generation has advanced rapidly, from symbolic modeling~\cite{huang2018music} and waveform synthesis~\cite{dhariwal2020jukebox} to high-quality text-conditioned generation~\cite{copet2023simple, agostinelli2023musiclm, liu2025songgen, gong2025ace}. Recent evaluations further consider multiple aspects including production quality, enjoyment, and audio-text alignment~\cite{tjandra2025meta, yao2025songeval}.


However, existing systems mainly optimize population-level quality or explicit controllability, while personalized preference alignment remains underexplored. Musical preferences are individual-specific and associated with personality, culture, and self-identity~\cite{rentfrow2003re, nave2018musical, greenberg2022universals}. The MUSIC model~\cite{rentfrow2011structure} and later meta-analyses~\cite{schafer2017can} further show that musical taste has stable structure across listeners. A track can therefore satisfy generic quality criteria while remaining poorly matched to the person who receives it.

PAG-Agent addresses this gap with an explicit Personal Aesthetic Profile. We use a profiling MLLM to distill favorite-song audio and contextual user information into a structured Personal Aesthetic Profile. A separate planning LLM translates this representation into track-level controls for a downstream music generator. This separates preference inference from music realization and keeps the user representation explicit.

The explicit representation also motivates a user-specific evaluation framework. PA-Eval treats successful personalization as requiring both consistency with the intended listener and specificity relative to other listeners. ACS measures alignment with the target user’s aesthetic reference, while ASI compares this alignment with cross-user matches, reducing the score when the same content also aligns well with other users.

Our contributions are summarized as follows. (1) We introduce PAG-Agent, which summarizes a user’s favorite songs and profile information into a Personal Aesthetic Profile and converts it into music-generation instructions in a zero-shot setting. (2) We propose PA-Eval, which measures both how well generated music matches the target user and how specific that match is compared to other users. (3) We build Personal-100, a dataset of 100 users, and evaluate PAG-Agent on multiple MLLM and music-generation backbones, along with a human study involving 50 participants.

\begin{figure*}[t]
  \centering
  \includegraphics[width=0.85\textwidth]{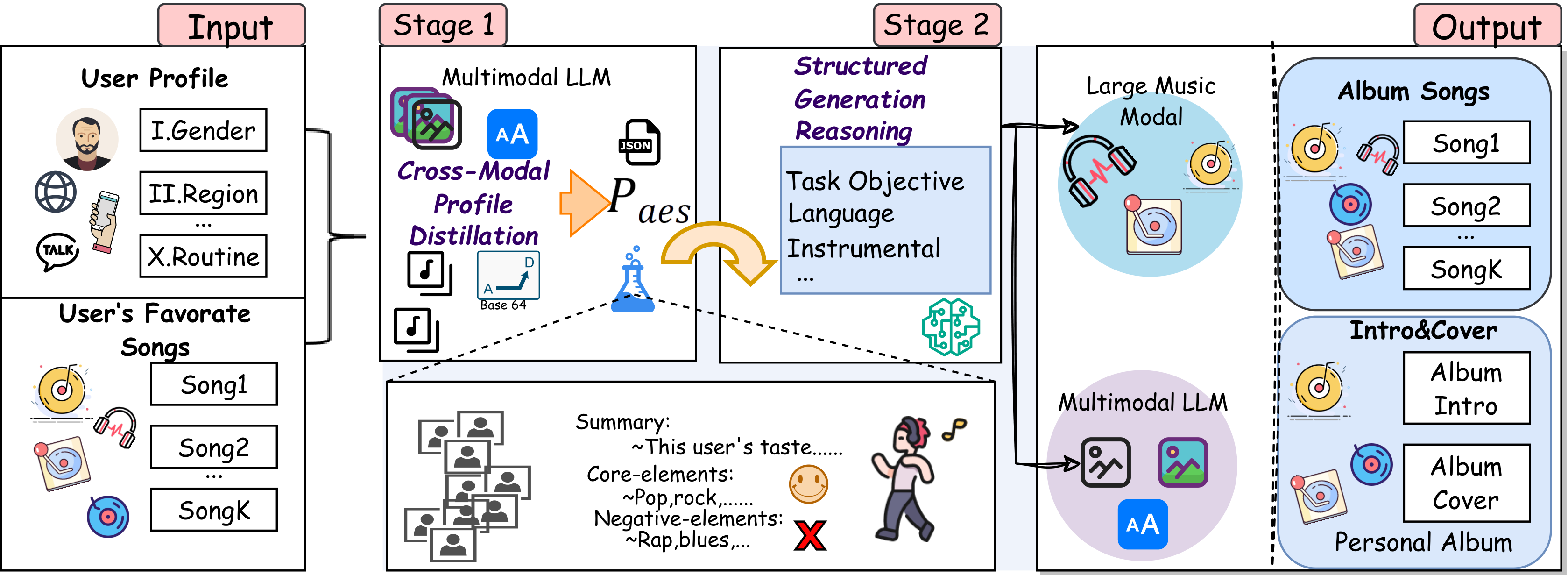}
  \caption{Overview of PAG-Agent. The profiling module distills favorite songs and contextual user information into a Personal Aesthetic Profile. The planning module converts that profile into track-level specifications for a music generator. The same profile can also condition cover art and a short album narrative, while the audio tracks remain the primary evaluation target in this work.}
  \label{fig:agent}
\end{figure*}

\section{Related Work}
\label{sec:related}

\textbf{Text-to-music generation.}

The field has progressed from symbolic sequence modeling~\cite{huang2018music} and waveform synthesis~\cite{dhariwal2020jukebox} to high-fidelity text-conditioned generation~\cite{copet2023simple,agostinelli2023musiclm,chen2024musicldm,gong2025ace}. Multimodal architectures further connect music understanding with generation~\cite{liu2023m}, while recent systems improve controllability and aesthetic alignment~\cite{melechovsky2024mustango,liu2025songgen}. Many of these systems expose control through explicit prompts or predefined attributes. PAG-Agent instead constructs an intermediate user-specific aesthetic representation from listening evidence and profile context before producing generation instructions.

\textbf{Musical aesthetics and personalization.}
Recent benchmarks evaluate production quality, enjoyment, and audio-text correspondence~\cite{tjandra2025meta, yao2025songeval}, while broader surveys discuss the difficulty of evaluating generative music~\cite{10.1145/3769106}. Such population-level measures remain valuable for audio quality but do not directly characterize individual preference alignment. Research in music psychology shows that musical preferences are associated with stable individual differences, including personality and cognitive style~\cite{rentfrow2003re,rentfrow2011structure,greenberg2015cognitive,flannery2021musical}. Preference patterns also vary with age, cultural background, and broader listener context~\cite{bonneville2013music,greenberg2022universals,veenstra2015class,schedl2018challenges}. Recent listener studies further show that perceived authorship, emotional resonance, and perceived humanness can shape judgments of AI-generated music~\cite{lecamwasam2025exploring,figueiredo2025echoes}. PAG-Agent builds on these findings to model listener-specific aesthetic context as a complement to generic quality evaluation.
\section{PAG-Agent}
\label{sec:method}

\subsection{Preference to Control Factorization}

For each user \(u\), a musical history \(S_u=[s_1,\ldots,s_K]\) provides favorite-song evidence, while a contextual profile \(P_u\) contains attributes such as age, occupation, MBTI type, cultural background, and social activity level. PAG-Agent first represents the user’s preferences and then translates them into generation controls.

\[
P_{\mathrm{aes}}=G_{\mathrm{prof}}(S_u,P_u),
\qquad
Z_u=G_{\mathrm{plan}}(P_{\mathrm{aes}}),
\]

followed by

\[
M_u=G_{\mathrm{music}}(Z_u),
\qquad
A_u=(I,M_u,T).
\]

Here \(P_{\mathrm{aes}}\) represents the user's inferred aesthetic preferences and \(Z_u\) expresses those preferences as generator-facing controls. The album \(A_u\) contains generated music \(M_u\), cover art \(I\), and a short narrative \(T\). Because user preferences are modeled separately from music generation, the same Personal Aesthetic Profile can be used with different music generators without retraining.

\subsection{Profile Distillation}

To make user preferences easier to translate into generation controls across different music generators, the Personal Aesthetic Profile summarizes the original user evidence into an explicit user-level representation. The profiling MLLM \(G_{\mathrm{prof}}\) integrates the raw audio in \(S_u\) with contextual information \(P_u\) and produces a JSON-formatted representation containing an aesthetic summary, core stylistic attributes, and negative constraints. These fields capture recurring information about genre, instrumentation, mood, and incompatible styles.

\subsection{Structured Generation}

The planning module \(G_{\mathrm{plan}}\) translates \(P_{\mathrm{aes}}\) into track-level controls for emotion, vocal form, genre, instrumentation, and style, with all tracks derived from the same profile but allowed to vary in their musical characteristics. The profile also conditions \(G_{\mathrm{media}}\) to generate cover art and a short album narrative, while the experiments focus on generated music as the primary evaluation target.

\section{PA-Eval}
\label{sec:eval}

\subsection{Personal-100 Dataset}

We constructed Personal-100 with 100 participants from multiple countries and occupations, following the ethical procedures described in Sec.~\ref{sec:eth}. Each participant provided a pool of ten favorite songs \(S_{\mathrm{pool}}\) and contextual and demographic profile information \(P_u\). In the main experiments, \(K{=}8\) songs are sampled from \(S_{\mathrm{pool}}\) for each user, with the choice of \(K\) examined in Sec.~\ref{sec:ablation}.

\subsection{Six-Dimensional Aesthetic Space}
\label{sec:space}

We represent personal aesthetics in a six-dimensional space \(\mathbb{V}\) informed by prior work on musical preference structure and affect~\cite{rentfrow2003re, rentfrow2011structure,russell1980circumplex,juslin2008emotional}. The dimensions are Emotional Valence (EV), Energy and Arousal (EA), Sonic Texture Complexity (STC), Musical Sophistication (MS), Novelty and Unconventionality (NU), and Cultural and Stylistic Rebellion (CSR). Each dimension is assigned one of five discrete levels together with a textual description.

\begin{figure*}[t]
  \centering
  \includegraphics[width=0.85\textwidth]{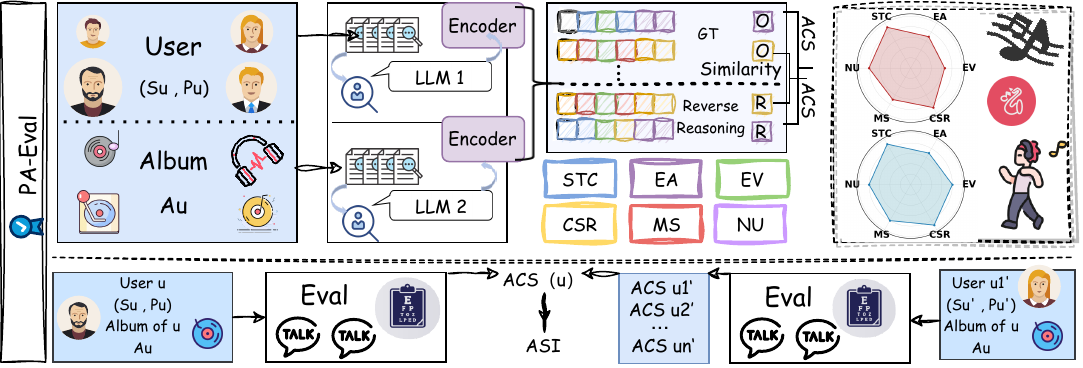}
  \caption{The PA-Eval pipeline. Independent reference and reverse streams map user evidence and generated output into a shared six-dimensional aesthetic space. ACS measures target-user consistency, while ASI compares that consistency against cross-user alignment.}
  \label{fig:eval}
\end{figure*}

Two independent MLLM streams map the user evidence and generated album into the same aesthetic space. \(E_{\mathrm{ref}}\), denoted as ``GT'' in Fig.~\ref{fig:eval}, derives the reference representation solely from \((S_u,P_u)\), whereas \(E_{\mathrm{rev}}\) derives the generated representation from \(A_u\) alone, without access to the user identity or profile. The dimension descriptions are then encoded by \(E_{\mathrm{text}}\) in \(\mathbf{V}_{\mathrm{ref}}\) and \(\mathbf{V}_{\mathrm{gen}}\), so the comparison depends only on the information inferred from the generated output on the generation side.

The follow-up human evaluation in Sec.~\ref{sec:human} provides independent evidence that ACS tracks subjective preference.

\subsection{Metrics}

PA-Eval evaluates personalization through complementary consistency and specificity. ACS measures alignment with the intended user, while ASI contrasts this alignment against other users to distinguish user-specific generation from broadly appealing content.

\textbf{Aesthetic Consistency Score.}
For aesthetic dimension \(i\), the level discrepancy is
\[
\Delta_i=\left|L_{\mathrm{ref}}^{(i)}-L_{\mathrm{gen}}^{(i)}\right|.
\]
We use a calibrated level-aware weight
\[
p_i(\Delta_i)
=
-\alpha\frac{e^{\Delta_i+\gamma}}{\Delta_i+\gamma}+\beta
\]
with \(\alpha{=}0.1\), \(\beta{=}e/10+1\), and \(\gamma{=}1\). This calibration gives unit weight under exact level agreement, retains substantial weight for small discrepancies, and reverses the contribution under large contradictions. ACS combines the weight with the corresponding aesthetic embeddings
\[
\mathrm{ACS}
=
\frac{
\sum_{i=1}^{6}
p_i(\Delta_i)
\left(
\mathbf{v}_{\mathrm{ref}}^{(i)}
\cdot
\mathbf{v}_{\mathrm{gen}}^{(i)}
\right)
}{
\sqrt{\sum_{i=1}^{6}\left(\mathbf{v}_{\mathrm{ref}}^{(i)}\right)^2}
\sqrt{\sum_{i=1}^{6}\left(\mathbf{v}_{\mathrm{gen}}^{(i)}\right)^2}
}.
\]

\textbf{Aesthetic Specificity Index.}
High target-user consistency can still arise from broadly appealing content. ASI therefore contrasts self-alignment with weighted cross-user alignment~\cite{hadsell2006dimensionality,forer1949fallacy}
\[
\mathrm{ASI}
=
\frac{1}{6}
\sum_{i=1}^{6}
\left[
C_{\mathrm{self}}^{(i)}
-
\frac{1}{N}
\sum_{u'\neq u}
w_i(u,u')C_{\mathrm{cross}}^{(i)}
\right].
\]
Here \(C_{\mathrm{self}}^{(i)}\) is the target user's per-dimension consistency term. \(C_{\mathrm{cross}}^{(i)}\) replaces the target reference with the reference representation of a comparison user. The weight \(w_i(u,u'){=}1-\mathrm{sim}^{(i)}(u,u')\) reduces the contribution of comparison users whose preferences are already similar to the target. Higher ASI indicates that the observed alignment is more specific to the intended user.

\section{Experiments}
\label{sec:exp}

\subsection{Setup}
\label{sec:setup}

The main evaluation uses Personal-100 with \(K{=}8\), while ablations use the Personal-25 subset. PAG-Agent uses Gemini-3.0-Pro or a GPT stack in which GPT-4o~\cite{hurst2024gpt} processes audio and GPT-5.2 performs structured planning. Music is generated with Suno-V5 or ACE-Step~\cite{gong2025ace}. PA-Eval uses Gemini-3.0-Pro or the same GPT stack for both evaluation streams, while OpenAI text-embedding-3-large provides \(E_{\mathrm{text}}\). AudioBox Aesthetics~\cite{tjandra2025meta} and CLAP~\cite{10095969} provide additional quality and alignment evaluation.

The \emph{Random} baseline uses a generic prompt that does not contain user information. The \emph{Direct} baseline receives the same songs and profile as PAG-Agent through a single instruction, then produces a generation prompt without constructing the intermediate Personal Aesthetic Profile. This comparison tests whether explicit profile distillation and structured planning improve over direct conditioning with the same user evidence.

\subsection{Main Results}

\begin{table}[t]
\centering
\small
\setlength{\tabcolsep}{3.2pt}
\begin{tabular}{lcc}
\toprule
\textbf{Method} & \textbf{ACS\(\uparrow\)} & \textbf{ASI\(\uparrow\)} \\
\midrule
\multicolumn{3}{l}{\textit{PAG MLLM Gemini \quad Eval MLLM Gemini \quad ACE-Step}} \\
Random & \(0.51{\pm}0.15\) & \(0.37{\pm}0.19\) \\
Direct & \(0.58{\pm}0.11\) & \(0.44{\pm}0.15\) \\
\textbf{PAG-Agent} & \(\mathbf{0.69{\pm}0.10}\) & \(\mathbf{0.56{\pm}0.14}\) \\
\midrule
\multicolumn{3}{l}{\textit{PAG MLLM Gemini \quad Eval MLLM Gemini \quad Suno-V5}} \\
Random & \(0.56{\pm}0.12\) & \(0.41{\pm}0.12\) \\
Direct & \(0.68{\pm}0.10\) & \(0.49{\pm}0.10\) \\
\textbf{PAG-Agent} & \(\mathbf{0.75{\pm}0.07}\) & \(\mathbf{0.63{\pm}0.06}\) \\
\midrule
\multicolumn{3}{l}{\textit{PAG MLLM GPT-4o/5.2 \quad Eval MLLM Gemini \quad Suno-V5}} \\
Random & \(0.55{\pm}0.10\) & \(0.35{\pm}0.13\) \\
Direct & \(0.67{\pm}0.11\) & \(0.41{\pm}0.12\) \\
\textbf{PAG-Agent} & \(\mathbf{0.73{\pm}0.06}\) & \(\mathbf{0.53{\pm}0.11}\) \\
\midrule
\multicolumn{3}{l}{\textit{PAG MLLM GPT-4o/5.2 \quad Eval MLLM GPT-4o/5.2 \quad Suno-V5}} \\
Random & \(0.52{\pm}0.12\) & \(0.39{\pm}0.13\) \\
Direct & \(0.64{\pm}0.18\) & \(0.45{\pm}0.12\) \\
\textbf{PAG-Agent} & \(\mathbf{0.71{\pm}0.11}\) & \(\mathbf{0.57{\pm}0.12}\) \\
\bottomrule
\end{tabular}
\caption{PA-Eval results on Personal-100. Values report mean \(\pm\) standard deviation across users. PAG-Agent improves both metrics across every displayed backbone configuration.}
\label{tab:main}
\end{table}

Across the four backbone configurations in Table~\ref{tab:main}, including changes to the generation MLLM, evaluation MLLM, and music generator, PAG-Agent improves ACS by roughly 9 to 19 percent and ASI by 27 to 29 percent over Direct. Compared with Random, the gains reach roughly 33 to 37 percent in ACS and 46 to 54 percent in ASI.

Generic audio quality remains comparable across methods. In the original AudioBox Aesthetics evaluation, Production Quality remains approximately 7.9 to 8.2 and Content Enjoyment remains approximately 7.3 to 7.7 across the tested settings. CLAP similarity also stays in a comparable range. These results suggest that the personalization gains are not explained by higher generic audio quality alone.

\subsection{Human Evaluation}
\label{sec:human}

We re-contacted 50 participants from Personal-100 for a follow-up evaluation. Participants rated generated albums on a 1 to 5 Likert scale for musicality and subjective preference. Musicality was defined independently of personal taste, and participants were not told that the evaluated content had been customized for them.

\begin{table}[t]
\centering
\small
\begin{tabular}{lcc}
\toprule
\textbf{Method} & \textbf{Musicality} & \textbf{Preference} \\
\midrule
Random     & 4.30 & 3.68 \\
Direct     & \textbf{4.42} & 4.12 \\
PAG-Agent  & 4.40 & \textbf{4.76} \\
\midrule
\(r_{\mathrm{ACS/Human}}\) & \(-0.16\) & \textbf{0.64} \\
\bottomrule
\end{tabular}
\caption{Mean human ratings from 50 participants. PAG-Agent receives the highest mean preference while musicality remains similar across methods. ACS is positively associated with subjective preference and shows little association with generic musicality.}
\label{tab:human}
\end{table}

Table~\ref{tab:human} shows similar mean musicality across all three methods. Mean subjective preference increases from 3.68 for Random and 4.12 for Direct to 4.76 for PAG-Agent. Across participants, ACS correlates positively with subjective preference (\(r{=}0.64\)) but only weakly with musicality (\(r{=}{-}0.16\)), indicating that ACS tracks personal preference alignment while showing little association with generic musicality.

\subsection{Ablation Studies}
\label{sec:ablation}

Ablations use Personal-25 with Suno-V5.


\begin{table}[t]
\centering
\small
\begin{tabular}{lcc}
\toprule
\textbf{Configuration} & \textbf{ACS} & \textbf{ASI} \\
\midrule
Songs only \(S_u\) & 0.60 & 0.42 \\
Profile only \(P_u\) & 0.61 & 0.47 \\
Songs + sparse \(P_u\) & 0.68 & 0.54 \\
\textbf{Songs + full \(P_u\)} & \textbf{0.74} & \textbf{0.62} \\
\bottomrule
\end{tabular}
\caption{Modality ablation with Gemini. Combining song history with the full profile gives the highest ACS and ASI.}
\label{tab:abl_mod}
\end{table}


Across \(K\in\{4,6,8,9,10\}\), the input-song ablation peaks at \(K{=}8\) for both Gemini (ACS/ASI = 0.74/0.62) and GPT (0.71/0.57). Neighboring \(K\) values remain close in ACS but show larger changes in ASI, so we use \(K{=}8\) in the main experiments.

Table~\ref{tab:abl_mod} shows that either modality alone performs below the combined setting. Adding a sparse user profile to the song history raises ACS from 0.60 to 0.68 and ASI from 0.42 to 0.54. The full profile reaches 0.74 ACS and 0.62 ASI, indicating that richer contextual metadata consistently benefit personalization.

\subsection{Backbone and Dimensional Analysis}
\label{sec:analysis}

Under the matched Gemini configuration in Table~\ref{tab:main}, Suno-V5 yields higher absolute ACS and ASI than ACE-Step across Random, Direct, and PAG-Agent, indicating a generator-dependent shift in attainable alignment. PAG-Agent remains above Direct with both generators, showing that the benefit of profile distillation is retained across synthesis models.

\begin{table}[t]
\centering
\small
\setlength{\tabcolsep}{2.5pt}
\begin{tabular}{lccc}
\toprule
\textbf{Dimension} & \textbf{ACS} & \textbf{ASI} & \(r_{\mathrm{ref,gen}}\) \\
\midrule
Emotional Valence (EV) & 0.68\(\pm\)0.12 & 0.49\(\pm\)0.12 & 0.72 \\
Energy \& Arousal (EA) & \textbf{0.79\(\pm\)0.04} & 0.64\(\pm\)0.07 & \textbf{0.76} \\
Sonic Texture (STC) & 0.77\(\pm\)0.05 & 0.62\(\pm\)0.09 & 0.75 \\
Musical Sophistication (MS) & 0.60\(\pm\)0.14 & 0.39\(\pm\)0.18 & 0.69 \\
Novelty (NU) & 0.70\(\pm\)0.08 & 0.57\(\pm\)0.10 & 0.73 \\
Cultural Rebellion (CSR) & \textbf{0.79\(\pm\)0.05} & \textbf{0.65\(\pm\)0.10} & \textbf{0.76} \\
\midrule
Average & 0.72\(\pm\)0.04 & 0.56\(\pm\)0.05 & 0.74 \\
\bottomrule
\end{tabular}
\caption{Per-dimension results with Gemini and Suno-V5. \(r_{\mathrm{ref,gen}}\) is computed across users within each dimension.}
\label{tab:dim}
\end{table}

Table~\ref{tab:dim} further shows variation across aesthetic dimensions. Energy \& Arousal and Cultural Rebellion achieve the highest ACS at 0.79, while Musical Sophistication obtains the lowest ACS and ASI at 0.60 and 0.39. Reference and generated scores remain positively correlated across every dimension, with \(r\) ranging from 0.69 to 0.76 and averaging 0.74, indicating that alignment extends across the aesthetic space instead of being concentrated in a single attribute.

\section{Conclusion}


We presented PAG-Agent, a zero-shot framework for personalized music album generation through explicit aesthetic profile distillation. PA-Eval evaluates both target-user alignment and cross-user specificity. Experiments on Personal-100 show consistent improvements over generic and direct conditioning, while human evaluation confirms that ACS correlates with subjective preference.

\section{Compliance with Ethical Standards}
\label{sec:eth}

This study was reviewed and approved by an Institutional Review Board (IRB). All participants in the Personal-100 dataset and the 50-participant follow-up evaluation provided documented informed consent acknowledging voluntary participation. Sensitive data including demographic metadata and favorite songs are used solely for academic research. All data were de-identified to the extent possible, and the raw dataset will not be publicly released to mitigate re-identification risks. Input songs are used strictly as acoustic anchors for style analysis rather than melodic reproduction, and prompts are designed to avoid soliciting specific copyrighted material. No funding was received for conducting this study. The authors have no relevant financial or nonfinancial interests to disclose.

\bibliographystyle{IEEEbib}
\bibliography{custom}

\end{document}